# Dark matter signatures in cosmic rays

M. Maroudas[1], E. Akman[2], A. Argiriou[3], G. Cantatore[4], E. Georgiopoulou[3], S. Hofmann[5], M. Karuza[6], A. Kryemadhi[7], I. Lazanu[8], A. Mastronikolis[9], M. Parvu[8], Y. K. Semertzidis[10], I. Tsagris[3,&], M. Tsagri[3,&], G. Tsiledakis[11], K. Zioutas[3,*]

[1] *Institute for Experimental Physics, University of Hamburg, 22761 Hamburg, Germany*
[2] *Department of Physics, University of California, Berkeley USA*
[3] *Physics Department, University of Patras, 26504 Patras-Rio, Greece*
[4] *Physics Department, University & INFN of Trieste, 34127 Trieste, Italy*
[5] *Independent Researcher, Fastlinger Strasse 17, 80999 München, Germany*
[6] *Physics Department, University of Rijeka, 51000 Rijeka, Croatia*
[7] *Physics Department, Messiah University, Mechanicsburg, PA 17055, USA*
[8] *University of Bucharest, Faculty of Physics, MG-11 Magurele-Bucharest, Romania*
[9] *Imperial College, SW7 2AZ London, United Kingdom*
[10] *KAIST, Daejeon, Korea*
[11] *IRFU/CEA-Saclay, CEDEX, 91191 Gif-sur-Yvette, France*
[&] *Present address: Geneva, Switzerland*
[*] *E-mail: zioutas@cern.ch*

## Abstract:

Solar system observables such as sunspots, solar flares etc, exhibit unexpected planetary dependencies, even though no remote force beyond gravity is known to act over such distances. A possible explanation arises from cosmological models predicting streams of dark matter (DM) particles. Such streams can be gravitationally focused by solar system bodies, including the Moon, creating focal regions within the solar system that mimic non-existent remote planetary forces. These effects become apparent when event timings are projected onto planetary heliocentric longitudes. Here, we present the first analysis of an 11-year dataset of measured relativistic cosmic rays (CRs), focusing on positrons and antiprotons, to search for planetary relationships as the new signature for the dark sector. Significant patterns are observed in the rigidity range 1.0-1.92 GV, resembling previously reported planetary dependencies in solar and terrestrial observations. Two complementary analysis codes were employed. Interpolated daily rates within 27-day bins provided the first evidence for planetary dependencies, while direct raw data analysis reinforced the result with significances well above 5σ. The "product spectrum" of five Mars orbits also proves to be a powerful method to highlight coherent planetary dependencies. These findings support the idea that DM streams can be gravitationally focused by planets and the Moon toward Earth, occasionally producing strong flux amplifications. The observed relativistic positrons and antiprotons may originate from massive DM constituents annihilating or interacting within the solar system, possibly in Earth's atmosphere. Our results extend previous studies to direct measurements of CRs and suggest a novel planetary dependency fitting-in the streaming DM scenario. This motivates further searches with other CR species, particularly energetic gamma rays, to support the proposed DM signature.

## 1. Introduction

Dark matter (DM) was first proposed by Zwicky (1933), who observed cosmic gravitational phenomena inconsistent with known physics. This remains one of modern astrophysics' greatest challenges, as the universe appears to be dominated by an unknown substance. The term "*dark matter*" implies a component that neither emits nor reflects light, carries no electric charge, and interacts only feebly with ordinary matter, rendering it invisible to conventional telescopes [1]. Yet, this definition can be misleading, since most direct detection efforts with highly sensitive instruments assume that DM particles do interact, albeit weakly, with standard model (SM) particles such as photons, electrons, or atomic nuclei, producing detectable signals. Interestingly, also recent studies hint that DM may interact with ordinary matter more "strongly" (= large cross section) than previously assumed [2,3]. One promising window into such interactions is offered by energetic cosmic rays (CRs), being secondaries from the parent DM. While CRs are often attributed to distant astrophysical sources like supernovae or pulsars, it has been proposed that some of the high energy (HE) CRs could instead be produced locally, via self-annihilation, or interactions involving heavy DM components within the solar system [4–7]. Secondary antimatter CRs, like positrons and antiprotons, could carry imprints of such nearby processes. This possibility raises a key question: could local gravitational structures, such as planets, moons and the Sun, modulate CR fluxes via their influence on DM streams? Antiprotons are particularly informative, since no primary astrophysical sources are expected, in contrast to positrons. Yet, the strongest signature identified in this study is found in cosmic positrons within the rigidity range 1.0–1.92 GV, while their flux decreases with higher rigidities.

The idea of the present work is rooted in gravitational focusing within the solar system. For slow-moving DM particles, the gravitational impact increases with $1/(\text{velocity})^2$, enhancing occasionally the local DM flux [10,11]. While the Sun is the strongest gravitational lens in the solar system, planets and even the Moon can also influence DM streams, particularly when they align with the orbital paths of solar system bodies, including their intrinsic mass distribution. This process is illustrated in Fig. 1, which depicts gravitational focusing of an invisible stream by the solar system. A 2012 study showed that planetary focusing could modulate DM fluxes intersecting Earth, potentially producing periodicities in solar and terrestrial observables [9,12-17]. The idea of planetary influence on solar activity itself dates to 1859, when Wolf noted the approximate match between the solar cycle and Jupiter's 11.86-year orbit [8]. Once dismissed as coincidence, recent findings suggest this may reflect an underlying physical mechanism by streaming DM [9].

Beyond solar effects, also several phenomena in Earth's atmosphere and its interior appear to correlate with planetary positions in ways not explained by standard models. These include anomalous stratospheric temperature excursions, changes in the ionospheric total electron content (TEC), and global seismic activity, including the clustering of large earthquakes (for magnitude M > 5.2) in certain planetary orbital positions [10-12]. Notably, TEC anomalies have been statistically linked to catastrophic earthquakes (M > 8), thus serving as precursors of about two months [12]. Simulations based on real atmospheric and seismic data suggest a potential for future earthquake warning, highlighting a societal spin-off following such investigations. A broader overview of planetary linked observables and gravitational focusing dynamics can be found in [10-13].

Within this context, we apply here for the first time two already established methodologies searching for planetary dependencies also in high energy CRs. If planetary focusing of DM streams can modulate atmospheric and seismic signals, they may also imprint even subtle periodicities in CR fluxes, e.g., in secondaries relativistic positrons and antiprotons [18-19]. These antimatter particles already exhibit unexplained spectral features, such as the positron excess above 10 GeV and a flattening in the antiproton spectrum, which are difficult to reconcile with standard models. Possible explanations include nearby pulsars or DM related processes [18-23]. Yet most approaches of CRs focus solely on their energy spectra. By contrast, we examine here potential relationships with planetary orbital phases, introducing a complementary probe of CR origins. If independently confirmed, such relationships would point to spatial or temporal structure in the CR flux not captured by conventional models, offering novel insight into the nature and distribution of initial DM components possibly contributing to the observed CRs. In this context, DM streams and clusters are instrumental. It is worth noticing that cosmological models naturally predict streams of well-motivated DM candidates, such as axions and WIMPs [16,17].

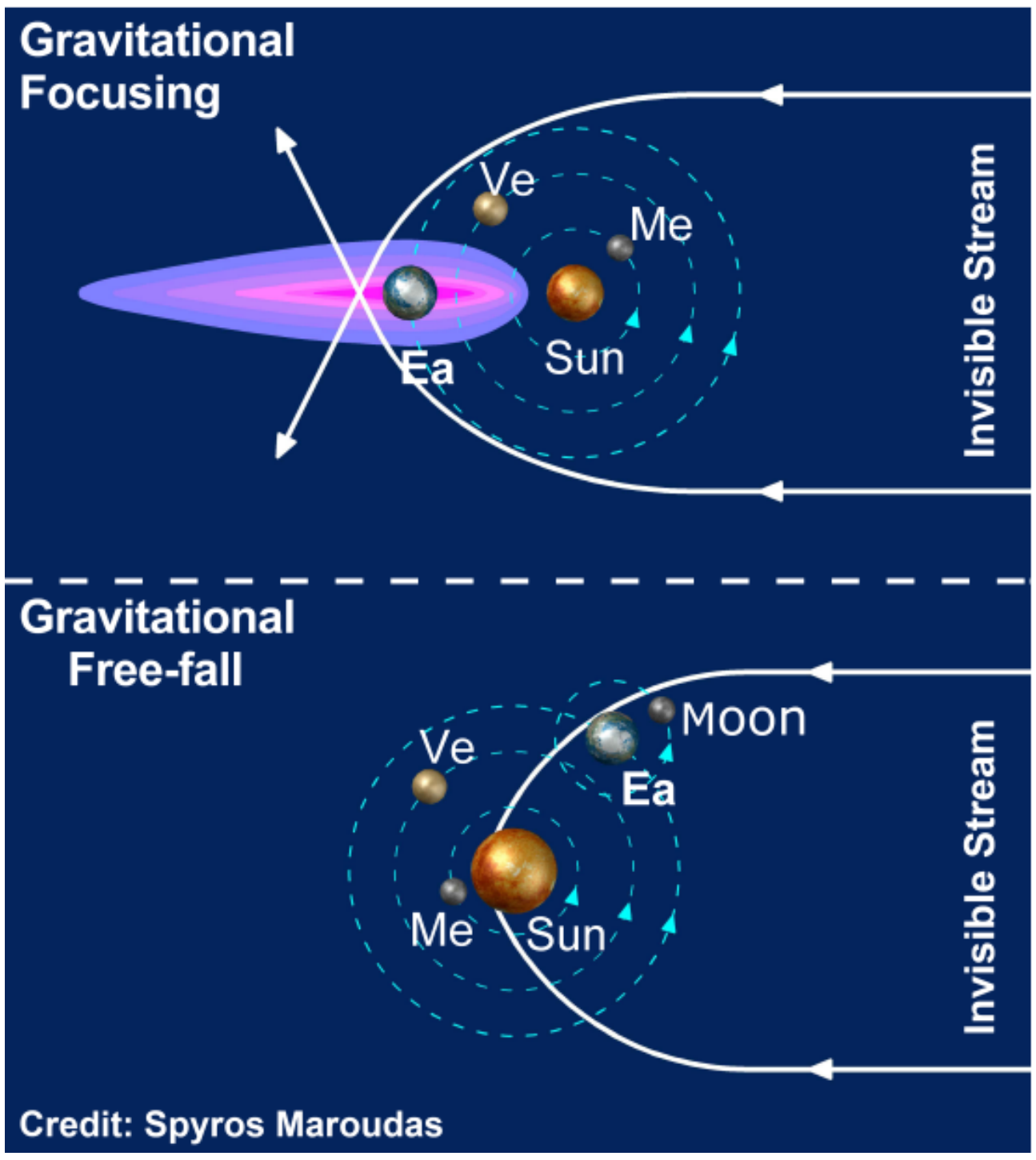

**Figure 1**: Graphic representation of gravitational focusing of an invisible stream by the solar system. (see ref. [10]).

## 2. Data origin and processing

The CR data analysed in this work originate from the alpha-magnetic spectrometer (AMS) aboard the International Space Station, as documented in Refs. [18–23]. The open availability of this HE dataset aligns with the broader Open Access initiative supported by CERN.

The measurements are provided as Bartels-rotation-binned time series (27-day intervals), grouped by rigidity. The full available observation period spans 11 years, from May 2011 to June 2022. While all available rigidity intervals were processed, our analysis emphasizes the lowest range (1.0–1.92 GV), where the flux is largest and planetary signatures are most clearly visible. At higher rigidities, the flux decreases rapidly, which may explain the increasingly random spectral shapes and the apparent absence of significant planetary dependencies, although physical effects cannot be excluded.

For each particle species, missing data points in the time series (Fig. 2) were interpolated linearly along the Bartels rotation axis. The number of the corrected interpolated points is small relative to the total number of bins: for positrons 6 out of 150 points (4%) and for antiprotons 11 out of 150 points (7.3%) were interpolated, respectively. The total uncertainty per data point was calculated by combining statistical, time-dependent, and systematic errors in quadrature. Each flux value is reported in units of [$m^{-2}$ $sr^{-1}$ $s^{-1}$ $GV^{-1}$], and all uncertainty components are also in this unit. As an example, the time series of positron and antiproton fluxes are shown in Fig. 2 for the 1.0–1.92 GV rigidity bin confirming that the limited interpolation does not distort the time structure of the underlying data.

Because of the coarse 27-day binning, as a first approximation, a linear interpolation between neighbouring bins is applied to derive daily flux values for positrons and antiprotons. This step provides a finer temporal structure and enables projection onto planetary orbital phases. The validity of this approximation has been tested previously[1]. After deriving daily values, the data are projected onto planetary heliocentric longitudes. The planetary positions were retrieved from NASA's Jet Propulsion Laboratory (JPL) Horizons System (https://ssd.jpl.nasa.gov/horizons.cgi) and matched to the CR flux data, allowing the construction of longitudinal distributions of the positron and antiproton fluxes projected onto planetary orbital positions. It is stressed, that due to orbital eccentricity, planets spend uneven time in different longitudes. We correct for this by normalizing the CR distributions with the time each planet spends in each longitude bin, removing thus geometric biases.

In addition, the raw AMS-02 measurements are analysed directly in their original 27-day bins. These data are projected onto planetary heliocentric longitudes without interpolation, providing a complementary approach that preserves the experimental uncertainties and allows proper statistical evaluation of modulations. Larger analysis bins are also applied to suppress binning artifacts while retaining reliable error estimates. Mars is particularly emphasized because of the relatively large number of bins (687/27 = 25).

---

[1] In a previous study (see [10] and Fig. 5 in ref. [4] therein), the number of measured daily solar flares over four solar cycles was first binned into monthly intervals and then back-converted to daily values by linear interpolation. Despite distortions in the spectral shape, the non-random longitudinal modulation remained clearly visible, revealing an underlying planetary relationship. This approach even allowed the study of data with larger initial binning (up to 72 days) [25]. This precedent validates applying a similar linear interpolation to CR data to first approximation of daily flux values. While precise statistical significance cannot be assigned to such reconstructed spectra, the recovery of a structured, non-random longitudinal distribution provides a first encouraging signal indication.

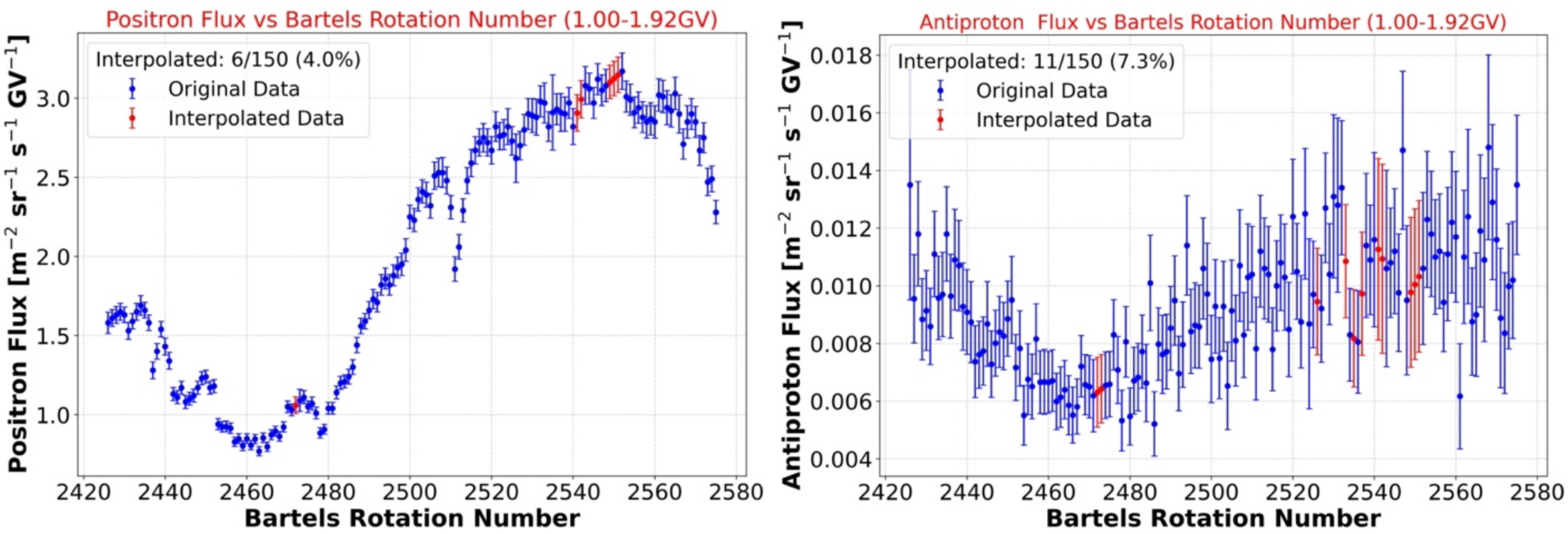

**Figure 2**: Time series of positrons (*left*) and antiprotons (*right*) flux for the 1.0–1.92 GV rigidity range, shown as a function of Bartels rotation number. The data collection period spans from 20 May 2011 to 24 May 2022. Red points indicate interpolated values. Error bars represent the total uncertainty per point, including statistical, time-dependent, and systematic contributions. The antiproton flux is more than 100 times smaller compared to positrons. Both distributions follow the 11 years solar cycle, which coincides with the synod of Venus-Earth-Jupiter-Saturn [9].

## 3. Data analysis and results

### A. Reconstructed daily fluxes

This work investigates whether the fluxes of HE positrons and / or antiprotons show significant planetary relationships, which would be a putative signature of streaming DM. Since relativistic particles are not affected by planetary gravitational impact within our solar system, such correlations, if observed, would point to another mechanism. One possibility is that these relativistic particles are secondaries produced from the self-annihilation, or interaction of massive slow-moving DM particles with normal matter [5-7], e.g. in atmospheres of the solar system, where they can escape from. In such a case, planetary gravitational focusing could modulate the flux of the slow-moving parent DM particles (v~300km/s), and this modulation would also be imprinted on the distribution of the resulting secondaries [5-7], namely relativistic positrons and antiprotons in this work. Therefore, while the secondary particles provide the time stamp of the signal, they are not necessarily the cause of the modulation.

An isotropic DM flux would produce a rather uniform CR signal, masking any planetary effects. In contrast, a DM component with a repeating planetary pattern, arising from a stream structure, could generate detectable asymmetries in the CR flux when analyzed in a planetary orbital frame. Preliminary spectral analysis using a Lomb-Scargle periodogram shows indications of peaks near 225 and 687 days, corresponding to the orbital periods of Venus and Mars, respectively (not shown). Building on prior experience with solar and terrestrial observables [10–12,24], we find that projecting event time stamps onto the concurrent planetary orbital positions provides a more sensitive method for detecting potential repeating planetary dependencies. This motivates the search for peaking distributions of CR positrons or antiprotons as a function of planetary heliocentric longitude. Mars and Venus are of particular interest because neither has a static magnetic field; for relativistic charged CRs, they effectively act as moving point-like masses, minimizing conventional magnetospheric interference. Over the 11-year observation period, Venus completes roughly 18 orbits, compared with five complete orbits of Mars.

In this framework, positron and antiproton fluxes are projected onto concurrent Venus' heliocentric longitude under different orbital conditions of Mars. Specifically, in Fig. 3, Mars' orbit is divided into two hemispheres (0°–180° vs. 180°–360°, and 270°–90° vs. 90°–270°), and the corresponding CR flux distributions are compared after being projected onto Venus' orbital position. If a modulated streaming DM flux interacts with the heliosphere in a way dependent on the Mars / Venus orbital configuration, the flux projected on Venus' orbit should also depend on Mars' position. Obviously, the observed flux can additionally depend on the positions of other planets due to their magnetic fields and combined heliospheric effects. The error bars in the projected flux distributions are linearly interpolated along with the 27-days flux values; they serve primarily for visual guidance, and the actual uncertainties at interpolated points may be larger. A more precise result, accounting for the known uncertainties, is presented in the following subsection.

Fig. 3 shows the projected fluxes for the rigidity interval 1.0–1.92 GV. Each row displays positron (top) and antiproton (bottom) distributions. The left column compares the distributions for Mars propagating in the sectors 270°–90°, 90°–270°, and no constraints. The right column shows results for 180°–360°, 0°–180°, and again the no-constraint reference (black curve in each plot). Deviations from the reference distribution are consistently visible, especially in the positron data. The stronger effects in certain Mars sectors suggest an orbital position dependency / influence, compatible with an external, directionally modulated source. More specifically, the structured differences observed in Fig. 3 fit-in a planetary-dependent modulation, potentially originating from a spatially anisotropic DM component. These asymmetries, especially in the positron channel, seem significant as they persist across four wide Mars-referenced orbital windows.

It is mentioned that the error bars in the projected flux distributions of Fig.3 are linearly interpolated along with the flux values; they are intended primarily for visual guidance. The actual uncertainties at the interpolated points may be larger, and therefore, the raw 27-day binned data are also analysed in the next subsection, to assess the statistical significance more robustly.

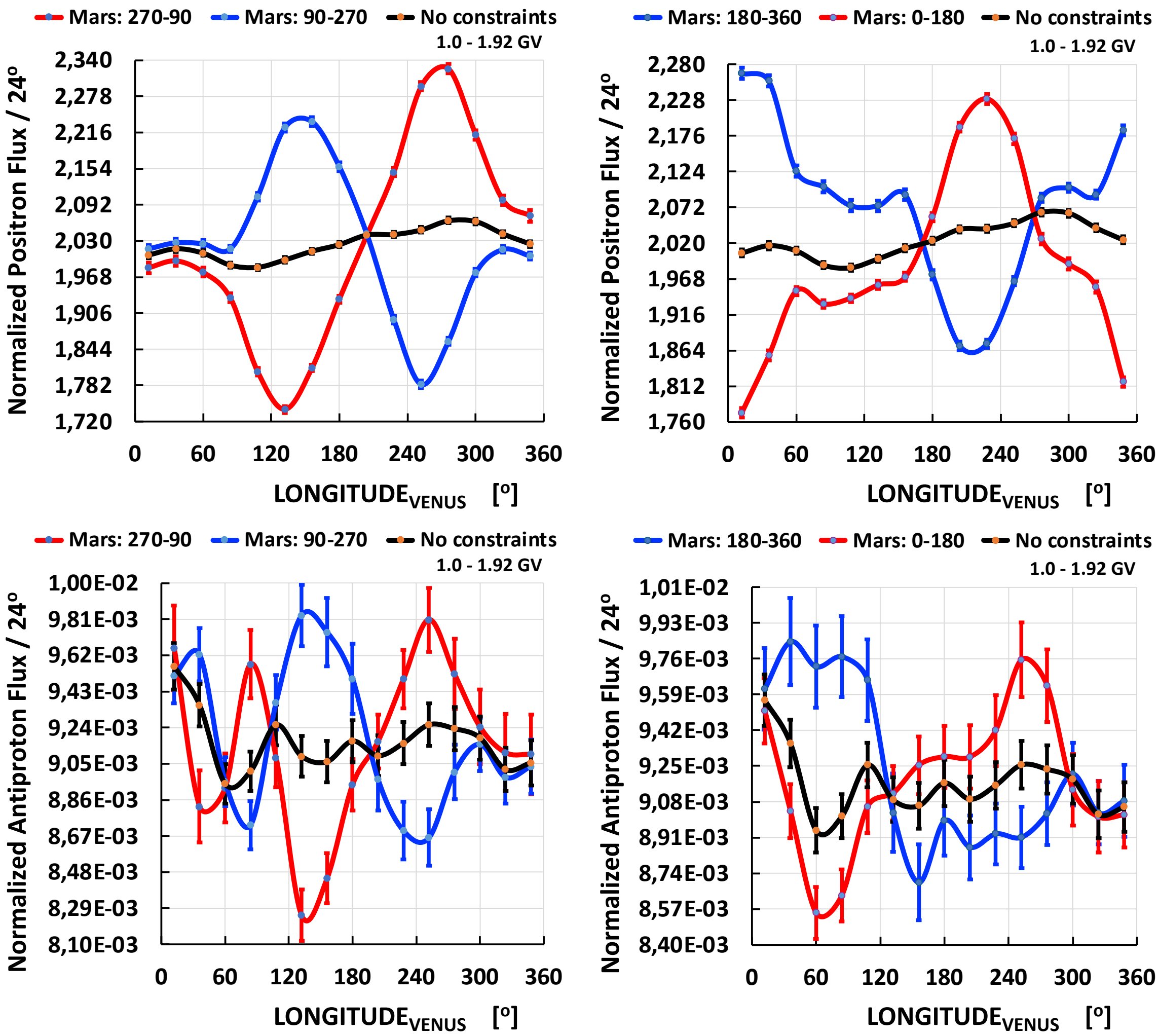

**Figure 3:** Time normalized fluxes of HE positrons (*top row*) and antiprotons (*bottom row*) projected onto Venus' heliocentric longitude, using a 24° binning and rigidity range 1.0–1.92 GV in all panels. Each subplot corresponds to a different 180°-wide Mars orbital sector. The black curve represents the reference case without any Mars constraint, also indicating an anisotropy and thus a possible planetary dependency. The potential indication of a planetary relationship is much larger for positrons compared to that of antiprotons. The data collection period spans from 20 May 2011 to 24 May 2022. Error bars are linearly interpolated between measurements for visual guidance; actual uncertainties may be larger.

## B. Raw data

The original AMS-02 positron data [18,22] are published in 27-day bins. To investigate possible planetary dependencies and search for statistically significant signatures of streaming dark matter (DM), we project the raw data onto the heliocentric longitude of Mars. Mars' orbital period (687 days) is much longer than the original binning, giving a ratio of 687/27 ≈ 25. This provides a large number of independent bins and thus a robust statistical basis. For comparison, Venus' orbital period corresponds to only ~8 bins (225/27), making Mars the more suitable reference frame.

The analysis uses exactly five complete Mars orbits (20 May 2011 – 8 October 2020). The end date is deliberately chosen to exclude any unfinished orbits, ensuring a consistent dataset. Time normalization is applied once more to account for Mars' orbital eccentricity, yielding flux distributions corrected for uneven planetary dwell times. Finally, for visualization, the data are rebinned into longitudinal intervals of 36° (~69 days), which balances statistical significance while minimizing binning effects.

The resulting time-normalized positron flux for the rigidity interval 1.0–1.92 GV as a function of Mars helicosentric position is shown in Fig. 4 left. The overall modulation amplitude between the minimum and maximum values is about 15%. More importantly, the third bin from the end, at 342° heliocentric longitude, representing a local maximum, exceeds the mean of the other nine bins with a statistical significance above 7σ, indicating that the observed behaviour is not a random fluctuation.

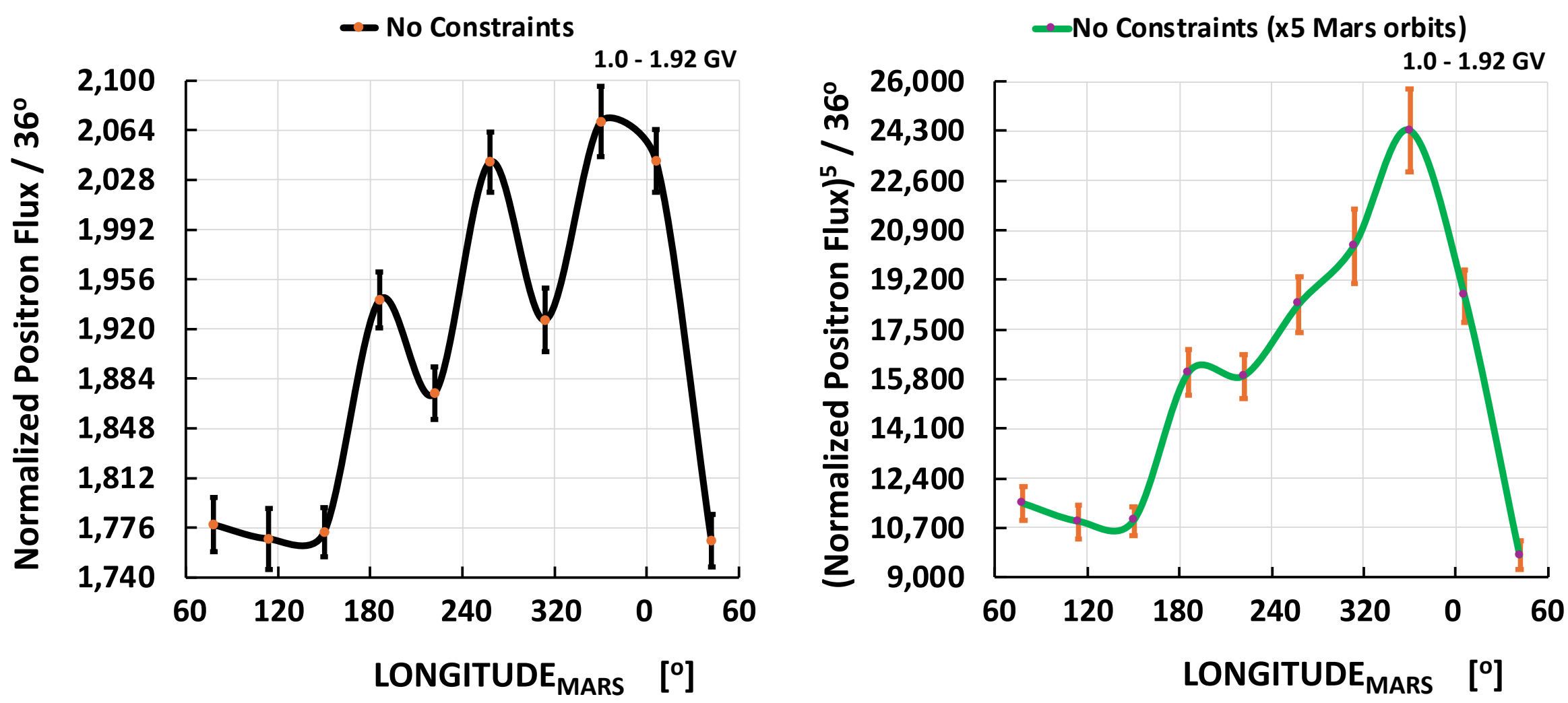

**Figure 4:** *Left*: Time-normalized positron flux (1.0–1.92 GV) projected onto Mars' heliocentric longitude from raw 27-day AMS-02 data, covering five complete Mars orbits (20 May 2011 – 8 Oct 2020). Longitudinal bin size: 36° (~69 days). The modulation amplitude is ~15%, with the 342° bin exceeding 7σ above the mean. *Right*: Product spectrum obtained by multiplying the five individual orbits bin by bin, which enhances coherent features. The errors in the product spectrum were propagated by adding the relative uncertainties of the individual orbits in quadrature. The resulting modulation amplitude reaches ~60%, with the maximum bin at 342° exceeding the mean by >6σ, consistent with a repeating Mars-orbital dependency.

In addition, to search for coherent planetary-dependent signals, the dataset is divided into five partial spectra, each corresponding to one complete Mars orbit (20 May 2011 – 25 Mar 2013, 25 Mar 2013 – 25 Feb 2015, 25 Feb 2015 – 31 Dec 2016, 31 Dec 2016 – 6 Nov 2018, and 6 Nov 2018 – 8 Oct 2020). Following previous work [24,25], these partial spectra (not shown here) are multiplied bin-by-bin to produce a "product spectrum", which is essential for the identification of streaming DM. This is because this method enhances features that persist coherently across all orbits, while reducing the relative proportion of uncorrelated noise. Errors in the product spectrum are propagated by adding the relative uncertainties of the individual orbits in quadrature, ensuring a consistent estimate of the statistical significance. This multiplication procedure reveals a much larger amplitude between maximum and minimum values (59.9%), highlighting repeating modulations consistent with a Mars orbital dependency (Fig. 4, right). Interestingly, the multiplied spectrum is also smoother than the sum spectrum, indicating that the amplification is due to coherent repeating signal(s) rather than statistical fluctuations, with the maximum bin at 342° exceeding the mean by >6σ.

This analysis demonstrates that the observed positron flux exhibits a coherent modulation over five consecutive complete Mars orbits. While relativistic positrons cannot be gravitationally focused by Mars, their production is consistent with secondary particles originating from streaming DM components experiencing planetary gravitational modulation within the solar system.

## 4. Discussion and conclusions

In this study, using the AMS-02 positron data, we find strong consistency with an underlying streaming DM scenario. This study investigates planetary dependencies in CRs, focusing on relativistic positrons in the low-rigidity range (1.0–1.92 GV), where fluxes are sufficient for statistical analysis. Higher-rigidity ranges show diminished modulation, likely due to limited statistics and/or underlying physical reasons. To our knowledge, this is the first analysis of CR measurements explicitly searching for planetary imprints as a novel signature of massive DM particles.

Previous studies reported planetary modulations with long-term solar and terrestrial observables, suggesting influence of a yet-unknown physical mechanism. The motivation for such investigations dates back to 1859, when Rudolf Wolf [8] noted the similarity between the 11-year solar cycle and Jupiter's 11.86-year orbital period, proposing that planetary influences might play a role in solar activity. More than 160 years later, the physical origin of the solar cycle remains unexplained within known physics despite the vast amount of data accumulated since then.

A 2013 study [9] proposed that planetary gravitational lensing of streaming DM could produce such modulations, provided the penetrating particles are slow enough to be focused within the solar system by planetary masses. Follow-up investigations using long-term datasets of solar and geophysical observables (see e.g. [10-13, 24-26] and references therein) supported the initial scenario [9]. The results suggested that a streaming, low-velocity component of DM might be identified through its secondaries and through its modulation periodicity, if it coincides with a single planetary or synodic periodicity.

The present work extends, for the first time, this line of research to CRs. Relativistic particles like positrons and antiprotons may originate from self-annihilation or interactions of massive, unseen particles such as DM. If such heavy parent particles exist and interact with solar system planetary gravitational potentials and SM particles, they may produce observable secondaries (e.g., positrons, antiprotons, etc.) with a characteristic planetary imprint. To test this, the time stamp provided by the observed CR events is projected onto the heliocentric orbital phase of the relevant planets, i.e., Venus and Mars allowing to examine whether the flux of positrons or antiprotons show systematic variations with planetary position. Interestingly, the observed non-random flux distributions are consistent with the presence of an anisotropic DM flux source population. Such a behaviour aligns with expectations of gravitationally focused DM streams rather than with the conventional isotropic CRs, lending confidence to the assumed streaming DM scenario.

Venus and Mars are the only planets relevant for these DM searches due to the absence of strong global magnetic fields. While the 27-day binning limits Venus analysis, the applied methodology can be extended to shorter-time datasets, allowing finer temporal resolution and potential detection of diurnal or other short-term periodicities. Combining analyses across planets may provide additional insights into the nature of the underlying DM streams.

Two complementary analysis methodologies were employed here. Fig. 3, based on reconstructed daily values from 27-day bins, provides initial evidence for a planetary dependency in positron flux providing a novel signature for streaming DM. Then Fig. 4 based on raw data, robustly confirms this first conclusion, particularly the Mars-orbital rhythm, and strengthens the case indicated by the interpolated analysis. The product spectrum plays a key role in this framework, as it amplifies coherent signals repeating across multiple orbits making it an instrumental tool for unravelling potential streaming DM signatures. Together, these methods demonstrate the robustness of the observed effect and illustrate the potential of finer binning to reveal sharper phase correlations and shorter-timescale structures.

Future studies should extend this approach to other CR species, including electrons and protons [20,21], which are primary CR components with different origins and propagation histories, as well as gamma rays measured by Fermi-LAT or other space and Earth-bound telescopes. Using finer binning, ideally daily or sub-daily, will be essential to improve sensitivity, resolve shorter periodicities, and assess transient effects. Identifying consistent planetary dependencies across species and datasets will provide a critical cross-check and further constrain DM-related interpretations.

In conclusion, our results provide strong evidence for a coherent Mars-orbit modulation of the positron flux, consistent with a streaming DM scenario. Relativistic positrons act as time-stamped tracers of gravitationally focused DM, likely streaming through the solar system. As higher-resolution CR datasets become available, the methodology presented here may allow the identification of DM constituents in CRs, potentially also revealing streams or clusters of axions or WIMPs of cosmic origin [16,17]. The combination of multiple CR species, finer temporal resolution, and independent datasets offers a promising path to establish planetary imprints as a robust signature of streaming DM.

## Acknowledgments

MM acknowledges funding by the Deutsche Forschungsgemeinschaft (DFG, German Research Foundation) under Germany's Excellence Strategy– EXC 2121 "Quantum Universe"– 390833306, and through the DFG funds for major instrumentation grant DFG INST 152/8241. This article is based upon work from COST Action COSMIC WISPers CA21106, supported by COST (European Cooperation in Science and Technology). KZ is grateful to the CAST collaboration at CERN, where the general concept behind this study was originally inspired in the context of efforts to enhance CAST's performance. This line of thought later gave rise to Refs. [9] and [10,26].

## Data Availability Statement

All data underlying this study are publicly available in references [18–23].

## Competing Interests

The authors have no competing interests to declare.

## Author Contributions

- Conceptualisation : KZ
- Investigation: KZ, MM
- Methodology: KZ, MM
- Data curation: MM
- Formal Analysis: MM, KZ
- Visualization: MM, IL, KZ
- Validation: ALL
- Writing and editing the manuscript: ALL

All authors have read and agreed to the published version of the manuscript.